\documentclass[a4paper,11pt]{elsarticle}
\pdfoutput=1 

\usepackage[T1]{fontenc} 

\newcommand{\R}{\mathbb R}

\newcommand{\Z}{\mathbb Z}

\newcommand{\dif}{\mbox{d}}
\usepackage{color}
\usepackage{url}
\def\vx{\ensuremath{\vec x}}
\def\vy{\ensuremath{\vec y}}

\usepackage[T1]{fontenc} 
\usepackage{amsmath, amsfonts, amssymb}
\usepackage{textgreek, stackengine, mathrsfs}
\usepackage{hyperref}
\usepackage{tikz}
\usetikzlibrary{decorations.pathmorphing}
\usepackage{pgfplots}
\usepackage[utf8]{inputenc}
\usepackage[toc,page]{appendix}

\usepackage{color}
\usepackage[normalem]{ulem}

\numberwithin{equation}{section}

\newcommand{\dk}{d\widetilde{k}\,}

\def\d{\ensuremath{\text{d}}}

\def\vx{\ensuremath{\vec x}}
\def\vy{\ensuremath{\vec y}}
\def\vz{\ensuremath{\vec z}}

\def\vN{\ensuremath{\vec\nabla}}

\def\vk{\ensuremath{\vec k}}
\def\vr{\ensuremath{\vec r}}
\def\mvx{\ensuremath{|\vx|}}

\begin{document}
\begin{frontmatter}
	
	\title{Polyharmonic Green Functions and Nonlocal BMS Transformations of a Free Scalar Field}
	
	\author{Carles Batlle\corref{cor1}\fnref{a}}
	\ead{carles.batlle@upc.edu}
	\author{V\'ictor Campello\fnref{b}}
	\ead{vicmancr@gmail.com}
	\author{Joaquim Gomis\fnref{c}}
	\ead{joaquim.gomis@ub.edu}
	
	\address[a]{Departament de Matem\`atiques and IOC, 
		Universitat Polit\`ecnica de Catalunya\\
		EPSEVG, Av. V. Balaguer 1, E-08800 Vilanova i la Geltr\'u, Spain}
	\address[b]{Departament de Matem\`atiques i Inform\`atica, Universitat de Barcelona, Gran Via de les Corts Catalanes 585, 
		E-08007, Barcelona, Spain}
	\address[c]{Departament de F\'isica Qu\`{a}ntica i Astrof\'isica and Institut de Ci\`{e}ncies del Cosmos (ICCUB), Universitat de Barcelona, Mart\'i i Franqu\`{e}s 1, E-08028 Barcelona, Spain
	}
	
	\cortext[cor1]{Corresponding author}

	\begin{abstract}
	We express the nonlocal BMS charges of a free massless Klein-Gordon scalar field in $2+1$ in terms of the Green functions of the polyharmonic operators. Using the properties of these Green functions, we are able to discuss the asymptotic behaviour of the fields that ensures the existence of the charges,  and prove that one obtains a realization of the $2+1$ BMS algebra in canonical phase space. We also discuss the transformations in configuration space, and show that in this case the algebra closes only up to skew-symmetric combinations of the equations of motion.
	 The formulation of the charges, in terms of Green functions, opens the way to the generalization of the formalism to other dimensions and systems.
	\end{abstract}

	
\end{frontmatter}

\flushbottom
\section{Introduction}

The BMS group of symmetry transformations \cite{Bondi:1962px,Sachs:1962wk} has experienced a surge of interest during the last decade. It has been used, for instance,  to deduce Weinberg's soft graviton theorems \cite{Weinberg:1965nx} as  Ward identities of 
BMS supertranslations \cite{He:2014laa,He:2014cra,Campiglia:2015qka,Kapec:2015ena}. 
A pedagogical overview of the role of BMS symmetries in several problems in field theory and gravitation is presented in \cite{Strominger:2017zoo}.
The  BMS algebra in the case of the  $2+1$ space-time has been studied in
\cite{Ashtekar:1996cd,Barnich:2006av}, and some applications can be found in \cite{Geiller:2021jmg}\cite{Fuentealba:2017omf}\cite{https://doi.org/10.48550/arxiv.1911.09651}\cite{Oblak:2015sea}\cite{Oblak:2016eij}.

Following the ideas in \cite{Longhi:1997zt}, an explicit nonlocal realization of supertranslations for scalar free fields in $2+1$ Minkowski space-time was developed in \cite{Batlle:2017llu}. The existence of the Noether charges  that canonically generate these transformations implies an asymptotic behaviour at spatial infinity for the scalar field.\footnote{The asymptotic behaviour of the scalar fields and the relation to soft theorems has been studied in
\cite{Campiglia:2017dpg}.}
The BMS asymptotic symmetries
at spatial infinity using the hamiltonian
formalism has been studied in
\cite{Henneaux:2018gfi,Henneaux:2018mgn,
Henneaux:2019yax}.

For a massless field, it was shown in \cite{Batlle:2017llu} that the functional variation of the field $\phi$ and its canonical momentum $\pi$ under a supertranslation is given by\footnote{The results for the massive case were also given in \cite{Batlle:2017llu}, as well as the extension to superrotations; see also \cite{Batlle:2017ghk}\cite{delmastro2017bms} for a discussion of the nonrelativistic limit, and \cite{Batlle2020} for the addition of super-dilatations.}
\begin{align}
	\delta_{\ell} \phi(t,\vec{x}) & = \int \d^2y\, [f_\ell(\vec{x}-\vec{y}) \phi(t,\vec{y}) + g_\ell(\vec{x}-\vec{y}) \pi(t,\vec{y})], \label{eq:deltaST_phi}\\
	\delta_{\ell} \pi(t,\vec{x}) & = \int \d^2y\, [h_\ell(\vec{x}-\vec{y}) \phi(t,\vec{y}) + f_\ell(\vec{x}-\vec{y}) \pi(t,\vec{y})]. \label{eq:deltaST_pi}
\end{align}
with $\ell\in\Z$  and where integration is all over two-dimensional space. The functions appearing in the above expressions are given by
\begin{align}
	f_\ell(\vec{x}) & = \quad 2\int \dk \omega\, \omega_\ell(\vec{k}) \sin(\vec{k}\cdot\vec{x}), \label{eq:fl}\\
	g_\ell(\vec{x}) & = \quad 2\int \dk \omega_\ell(\vec{k}) \cos(\vec{k}\cdot\vec{x}), \label{eq:gl}\\
	h_\ell(\vec{x}) & = -  2\int \dk \omega^2 \omega_\ell(\vec{k}) \cos(\vec{k}\cdot\vec{x}), \label{eq:hl}
\end{align}
with $\omega = |\vec{k}|$ and $\omega_\ell(\vec{k}) = \omega^{1-\ell}
(k_1+ik_2)^\ell$, and where the measure in momenta space is $\dk=\d^2k/((2\pi)^2 2\omega)$.
One can check that $\omega_{-\ell}=\omega_\ell^*$ and thus one can work with $\ell\geq 0$ and take the complex conjugate when negative indexes are needed. Furthermore, since $\omega_\ell(-\vk) =(-1)^\ell\omega_\ell(\vk)$, one can see from (\ref{eq:fl})(\ref{eq:gl})(\ref{eq:hl}) that
\begin{equation}
	f_{2\ell}(\vx) = 0,\quad  g_{2\ell+1}(\vx)=0,\quad h_{2\ell+1}(\vx)=0,\ \ \ell\in\Z.
	\label{eq:fgl0}
\end{equation}

Notice that, in general, the transformations (\ref{eq:deltaST_phi})(\ref{eq:deltaST_pi}) are nonlocal unless the functions $f_\ell$, $g_\ell$, $h_\ell$ are proportional to a delta function or a finite number of 
 its derivatives. As we will see, this happens only for $\ell=0,\pm 1$, which corresponds to ordinary space-time translations.

Using standard equal-time Poisson brackets and the properties $f_\ell(-\vec{x})=-f_\ell(\vec{x})$, $g_\ell(-\vec{x})=g_\ell(\vec{x})$, $h_\ell(\vec{x})=\nabla^2g_\ell(\vec{x})$, it can be seen that
the field transformations (\ref{eq:deltaST_phi}), (\ref{eq:deltaST_pi}) are generated by the supertranslation charges
\begin{align}
	Q_\ell(t) = \int \d^2x \d^2y\,  (f_\ell (\vec{x}-\vec{y})\pi(t,\vec{x}) \phi(t,\vec{y}) + \dfrac{1}{2} g_\ell (\vec{x}-\vec{y}) \pi(t,\vec{x}) \pi(t,\vec{y}) 
	 -\dfrac{1}{2} h_\ell (\vec{x}-\vec{y}) \phi(t,\vec{x}) \phi(t,\vec{y})). \label{eq:charge_st}
\end{align}
Once the asymptotic behaviour of $\phi(t,\vec{x}), \pi(t,\vec{x})$ at spatial infinity is given, see \ref{appG}, and using standard equal-time Poisson brackets, it can be shown (see \ref{appA}) that these charges have zero Poisson bracket with the Hamiltonian of the massless scalar field
\begin{equation}
H(t)=\int \d^2x \left(\frac{1}{2} \pi^2(t,\vec{x}) + \frac{1}{2} (\vec{\nabla}\phi(t,\vec{x}))^2\right),
\label{eq:H}
\end{equation}
and are thus conserved. The proof relies solely on the symmetry properties of the functions $f_\ell$ and $g_\ell$ and on the relation between $g_\ell$ and $h_\ell$.  The main goal of the paper is to show that the functions (\ref{eq:fl}), (\ref{eq:gl}) and (\ref{eq:hl}) can be cast in terms of higher level objects, which turn out to be Green functions, and to use their properties to discuss some aspects of the transformations. The algebra of  the transformations in terms only of 
$\phi(t,\vec{x})$ closes only up to an antisymmetric combination of the equations of motion.

The rest of the paper is organized as follows. Section \ref{sec2} presents
 the expression of the supercharges in terms of Green functions of appropriate operators. Using these, the existence of the charges is discussed. Section \ref{BMS-PS} computes the Poisson brackets of the obtained charges, while Section \ref{BMS-CS} considers the BMS transformations in configuration space. Finally, in Section \ref{diss} we summarize our results  and suggest that the formalism presented in this paper can be generalized to other dimensions and systems. Detailed calculations of all the results have been moved to the appendixes.

\section{Nonlocal transformation of the fields in terms of polyharmonic functions}
\label{sec2}
Using the explicit expression for $\omega_\ell$, one can write $g_\ell$ as
\begin{align}
	g_\ell(\vec{x}) & = \dfrac{1}{(2\pi)^2} \int d^2k\, \omega^{-\ell} (k^1+ik^2)^\ell \cos(\vec{k}\cdot\vec{x}). \label{eq:ST_g}
\end{align}
For $\ell$ odd, the integrand is antisymmetric in $\vec{k}$ and the integral
cancels out. For $\ell$ even and non-negative, one can write
\begin{align}
	g_{2\ell}(\vec{x}) & = \dfrac{1}{(2\pi)^2} \int d^2k\, \omega^{-2\ell} (k^1+ik^2)^{2\ell} \cos(\vec{k}\cdot\vec{x}) \nonumber\\
	& = (\partial_{x_1} + i \partial_{x_2})^{2\ell} (-1)^{\ell} \dfrac{1}{(2\pi)^2} \int d^2k\, \omega^{-2\ell} \cos(\vec{k}\cdot\vec{x}). \label{eq:g2l}
\end{align}
Now, defining a distribution $G_\ell (\vec{x})$ such that
\begin{equation}
	G_\ell (\vec{x}) = (-1)^\ell \dfrac{1}{(2\pi)^2} \int d^2k\, \omega^{-2\ell} \cos(\vec{k}\cdot\vec{x}),
\end{equation}
one gets the equation for a polyharmonic Green function,
\begin{equation}
	(\nabla_{\vec{x}}^2)^\ell G_\ell(\vec{x}) = \dfrac{1}{(2\pi)^2} \int d^2k\, \cos(\vec{k}\cdot\vec{x}) = \delta(\vec{x}), \quad \ell\geq 0.
	\label{eq:defGl}
\end{equation}
One has that $G_0(\vec{x})=\delta(\vec{x})$ and, for $\ell\geq 1$, the Green function is \cite{Cohl2012,Boyling1996}
\begin{equation} \label{eq:fundamental}
	G_\ell(\vec{x},\vec{y}) = G_\ell(\vec{x}-\vec{y}) =
	\dfrac{|\vec{x}-\vec{y}|^{2(\ell-1)}}{[(\ell-1)!]^2 2^{2\ell-1} \pi}(\log |\vec{x}-\vec{y}| - H_{\ell-1}),
\end{equation}
where $H_\ell = \sum_{i=1}^{\ell} \frac{1}{i}$ and $H_0 = 0$. It follows from (\ref{eq:defGl}) that 
\begin{equation}
\nabla^2 G_\ell(\vec{x}-\vec{y}) = G_{\ell-1}(\vec{x}-\vec{y}),
\quad \ell\geq1.
\label{eq:recurG}
\end{equation}
As shown in \ref{appC}, these functions satisfy also the convolution property
\begin{equation}
	\int\d^2x G_\ell(\vy-\vx) G_m(\vz-\vx) = G_{\ell+m}(\vy-\vz).
	\label{convGG}
\end{equation}

The expressions for $f_\ell$ and $h_\ell$ can then be directly obtained as a
function of $g_\ell$ by observing that $f_{2\ell+1} = -(\partial_{x_1} + i
\partial_{x_2}) g_{2\ell}$ and $h_{2\ell} = \nabla^2 g_{2\ell}$, for
$\ell\in\mathbb{N}$ and they are zero otherwise. Thus, these functions can be
written in terms of the polyharmonic Green function $G_\ell(\vec{x})$ as
\begin{align}
	g_{2\ell}(\vec{x}) & = (\partial_{x_1} + i \partial_{x_2})^{2\ell} G_\ell(\vec{x}), \label{eq:exp_gl}\\
	f_{2\ell+1}(\vec{x}) & = - (\partial_{x_1} + i \partial_{x_2})^{2\ell+1} G_\ell (\vec{x}), \label{eq:exp_fl}\\
	h_{2\ell}(\vec{x}) & = (\partial_{x_1} - i \partial_{x_2}) (\partial_{x_1} + i \partial_{x_2})^{2\ell+1} G_\ell(\vec{x}), \label{eq:exp_hl}
\end{align}
for $\ell\geq 0$. For $\ell=0$ one has $g_0(\vec{x})=\delta(\vec{x})$, $f_1(\vec{x}) = -(\partial_{x_1} + i \partial_{x_2})\delta(\vec{x})$ and $h_0(\vec{x})=\nabla^2\delta(\vec{x})$, which yield the standard space-time translations for the fields.

In terms of the $G_\ell$ the supertranslation charges (\ref{eq:charge_st}) for $\ell\geq 0$  take  the
forms
\begin{align}
	Q_{2\ell}(t) &= \int \d^2x\, \d^2y\, \left(  \dfrac{1}{2} \pi(t,\vec{x}) \pi(t,\vec{y})
	+ \dfrac{1}{2} \vec{\nabla} \phi(t,\vec{x}) \cdot \vec{\nabla} \phi(t,\vec{y}) \right) (\partial_{x_1} + i\partial_{x_2})^{2\ell} G_\ell (\vec{x}-\vec{y}), \label{eq:charge_st_polyharmonic_2l} \\
	Q_{2\ell+1}(t) &= \int \d^2x\, \d^2y\, (\partial_{x_1}+i\partial_{x_2})\pi(t,\vec{x})\phi(t,\vec{y}) (\partial_{x_1}+i\partial_{x_2})^{2\ell}G_\ell (\vec{x}-\vec{y})\label{eq:charge_st_polyharmonic_2l+1}\\
	      &=  \int \d^2x\, \d^2y\, \phi(t,\vec{x}) \pi(t,\vec{y}) (\partial_{x_1}+i\partial_{x_2})^{2\ell+1}G_\ell (\vec{x}-\vec{y})\\
	      &= - \int \d^2x\, \d^2y\, (\partial_{x_1}+i\partial_{x_2})\phi(t,\vec{x}) \pi(t,\vec{y}) (\partial_{x_1}+i\partial_{x_2})^{2\ell}G_\ell (\vec{x}-\vec{y}).\label{eq:2l+1}
\end{align}
Using that $\omega_\ell^*=\omega_{-\ell}$ one has that for $\ell<0$ the only difference is the appearance of $\partial_{x_1}-i\partial_{x_2}$ instead of $\partial_{x_1}+i\partial_{x_2}$, and thus
\begin{equation}
	Q_{-2\ell}(t) = Q_{2\ell}^*(t),\quad Q_{-(2\ell+1)}(t) = Q_{2\ell+1}^*(t).
	\label{ellneg}
\end{equation}

The supertranslation transformations in terms of the $G_\ell$ are given by
\begin{align}
	\delta_{2\ell}\phi(t,\vx) &= \{\phi(t,\vx),Q_{2\ell}(t)\} = \int\d^2y\,\pi(t,\vy) (\partial_{x_1} + i\partial_{x_2})^{2\ell} G_\ell (\vec{x}-\vec{y}),\label{STphi2l}\\
    \delta_{2\ell+1}\phi(t,\vx) &= \{\phi(t,\vx),Q_{2\ell+1}(t)\} = \int\d^2z\, \phi(t,\vz)  (\partial_{z_1}+i\partial_{z_2})^{2\ell+1}G_\ell (\vec{z}-\vec{x})\nonumber \\
    &= -\int\d^2y\, \phi(t,\vy)  (\partial_{x_1}+i\partial_{x_2})^{2\ell+1}G_\ell (\vec{x}-\vec{y}),\label{STphi2l+1}
\end{align}
where we have used the Poisson bracket $\{\phi(t,\vx),\pi(t,\vy)\}=\delta(\vx-\vy)$.

In particular, using that $G_0(\vx-\vy)=\delta(\vx-\vy)$, one can see that $Q_0(t)=H(t)$
is the generator of time-translations and that
\begin{align}
	Q_{x_1}(t) &= \frac{Q_1(t)+Q_{-1}(t)}{2} = -\int\d^2x\,  \pi(t,\vec{x}) \partial_{x_1}\phi(t,\vec{x}),\\
	Q_{x_2}(t) &= \frac{Q_1(t)-Q_{-1}(t)}{2i} = -\int\d^2x\,  \pi(t,\vec{x}) \partial_{x_2}\phi(t,\vec{x}),
\end{align}
generate the spatial translations.

We discuss next the asymptotic behaviour of the fields that guarantees the existence of the supertranslation charges and of the symplectic form.

Let us first consider the kinetic term in the action, which eventually leads to a well-defined Poisson bracket,
\begin{equation}
	\int\dif^2 x\, \dot\phi(t,\vx)\pi(t,\vx).
\end{equation}
If we assume asymptotic expansions
\begin{align}
\phi(t,\vx) &= \frac{\bar{\phi}_1}{\mvx} + \frac{\bar{\phi}_2}{\mvx^2} +  \ldots,\label{asphi}\\
\pi(t,\vx) &= \frac{\bar{\pi}_2}{\mvx^2} +   \frac{\bar{\pi}_3}{\mvx^3}+ \ldots,\label{aspi}
\end{align} 
where the $\bar{\phi}_1$, $\bar{\phi}_2$, $\bar{\pi}_1$, $\bar{\pi}_2$, \ldots, are functions depending on time and the angular variable then
\begin{equation}
	\int\dif^2 x\, \dot\phi(t,\vx)\pi(t,\vx) = \int\dif\theta\int r\dif r   \left(
	\dot{\bar{\phi}}_1  \bar{\pi}_2\frac{1}{r^3} + O(r^{-4})
	\right)
	\label{sf}
\end{equation}
which makes the term well defined. It follows also from these conditions that the field configuration has then a finite energy, and in fact the conditions cannot be relaxed, by instance by assuming $\phi\sim\log r$ or $\pi\sim 1/r$, if one wants to have a finite energy.  Notice that under these conditions  no logarithmic divergence appears  in (\ref{sf}), in contrast with the case in $3+1$ space-time discussed in \cite{Henneaux_2019}. 

The leading order behaviour of $G_l(\vx-\vy)$ for large $r=|\vx-\vy|$ is
\begin{equation}
G_\ell(r) \sim r^{2(\ell-1)}\log r.
\end{equation}
As shown in \ref{appG}, the derivatives of order $2\ell$ which appear in (\ref{eq:charge_st_polyharmonic_2l}) and (\ref{eq:2l+1}) behave as
\begin{equation}
(\partial_{x_1}+i\partial_{x_2})^{2\ell} G_\ell(r) \sim \frac{1}{r^2} \quad \forall \ell\geq 1.
\end{equation}
Taking this into account and comparing (\ref{eq:H}) with (\ref{eq:charge_st_polyharmonic_2l}) and (\ref{eq:2l+1}), it follows that the supertranslation charges exist for field configurations behaving as in (\ref{asphi}), (\ref{aspi}).

In order to have the BMS algebra we will need,
besides the generators of super-translations,  those of the Lorentz symmetries, given by
\begin{align}
	M^{12}(t) &= -\int\dif^2x\, \pi(t,\vx)\left(x_1\partial_{x_2}\phi(t,\vx)-x_2\partial_{x_1}\phi(t,\vx)\right)\\
	M^{0i}(t) &= -\int\dif^2x\, \left(t \pi(t,\vx)\partial_{x_i}\phi(t,\vx) +x_i {\cal H}(t,\vx)  \right),\quad i=1,2,
\end{align}
with 
\begin{equation}
	{\cal H}(t,\vx)= \frac{1}{2}(\pi^2(t,\vx)+ (\vec{\nabla}\phi(t,\vx))^2)
\end{equation}
the energy density of the scalar field.

\section{The BMS algebra in phase space}
\label{BMS-PS}
The abstract BMS algebra in $2+1$ is given by
\begin{align}
	[L_n,P_m] &= (n-m) P_{m+n},\\
	[P_m,P_{m'}] &=0,
\end{align}
with $n\in\{-1,0,1\}$ and $m,m'\in\Z$, and with the $L_n$ satisfying $[L_n,L_{n'}]=(n-n')L_{n+n'}$ and yielding the  $2+1$ Lorentz algebra. The algebra can be extended to $n,n'\in\Z$ by introducing the superrotations  $L_n$, $|n|>1$  \cite{Barnich:2011ct}\cite{Barnich:2011mi}.

 We will show that the above supertranslation charges (\ref{eq:charge_st_polyharmonic_2l}), (\ref{eq:charge_st_polyharmonic_2l+1}), (\ref{ellneg}) provide a realization of this algebra in terms of the standard equal-time  Poisson brackets and with the following combinations of Lorentz generators:
\begin{align}
	L_0(t) &= \frac{1}{2i}M^{12}(t),\\
	L_1(t) &= -M^{01}(t)-i M^{02}(t),\\
	L_{-1}(t) &= M^{01}(t)-i M^{02}(t) = -(L_1(t))^*.
\end{align}
The proof relies on the general symmetry properties of the Green functions $G_\ell$ and their derivatives, as well as on a key identity that is proved in  \ref{appB}. The brackets between the supertranslation charges are discussed in \ref{appA} and here we will discuss only those involving the Lorentz generators.

Let us consider first the Poisson bracket $\{L_0(t),Q_{2\ell}(t)\}$, for $\ell\geq 0$. With the notation $\phi(x)=\phi(t,\vx)$ and so on, and defining
\begin{equation}
	{\cal H}(x,y) =\frac{1}{2} \big(\pi(x)\pi(y)+\vN\phi(x)\cdot\vN\phi(y)\big),
\end{equation}
one has
\begin{eqnarray*}
\lefteqn{\{L_0(t),Q_{2\ell}(t)\}= } \\
&= -\frac{1}{2i}\int\d^2 x\d^2y\d^2z  \left\{
\pi(x) (x_1\partial_{x_2}\phi(x)-x_2\partial_{x_1}\phi(x)), {\cal H}(y,z)
\right\} (\partial_{y_1}+i\partial_{y_2})^{2\ell} G_\ell(\vy-\vz).
\end{eqnarray*}

This bracket is computed in \ref{appP}, and the result is (see (\ref{appPL2l}))
\begin{align}
	\{L_0(t),Q_{2\ell}(t)\} &= -2\ell Q_{2\ell}(t),\quad \ell\geq 0.
	\label{R2l}
\end{align}

Using similar steps to those of \ref{appP}, one can also obtain
\begin{align}
	\{L_0(t),Q_{2\ell+1}(t)\}  
	&= -(2\ell+1) Q_{2\ell+1}(t).\quad \ell\geq 0.
	\label{R2l+1}
\end{align}
The above computations are only valid for $\ell\geq 0$. For $\ell<0$ one can take the complex conjugate of (\ref{R2l}) and (\ref{R2l+1}), and use (\ref{ellneg}) and also $L_0^*(t)=-L_0(t)$. In this way one obtains, for $\ell\geq 0$,
\begin{align}
\{L_0(t),Q_{-2\ell}(t)\} &= 2\ell Q_{-2\ell}(t) = - (-2\ell) Q_{-2\ell}(t),\label{Rm2l}\\
\{L_0(t),Q_{-(2\ell+1)}(t)\} &= (2\ell+1) Q_{-(2\ell+1)}(t) = - (-(2\ell+1)) Q_{-(2\ell+1)}(t),\label{Rm2l+1}.
\end{align}
Relations (\ref{R2l}---\ref{Rm2l+1}) give the complete set of BMS algebra relations involving the rotation generator $L_0$.

Let us proceed now with the brackets involving the boost generators. Consider first
\begin{eqnarray*}
	\lefteqn{\{L_1(t),Q_{2\ell}(t)\}= } \\
	&=& \int\d^2 x\d^2y\d^2z  \Big\{
	t\pi(x) (\partial_{x_1}+i\partial_{x_2}) \phi(x) + (x_1+ix_2){\cal H}(x), {\cal H}(y,z)
	\Big\} (\partial_{y_1}+i\partial_{y_2})^{2\ell} G_\ell(\vy-\vz).
\end{eqnarray*}

As shown in \ref{appP}, this can be seen to be (see equation (\ref{appPB2l}))
\begin{align}
	\{L_1(t),Q_{2\ell}(t)\} &=(1-2\ell) Q_{2\ell+1}(t),
 \label{B2l}
\end{align}
which is the correct action of $L_1$ on a supertranslation of order $2\ell$, $\ell>0$, and can be extended to the trivial (Poincar\'e) case $\ell=0$.

Using similar computations, together with $G_\ell = \vN^2 G_{\ell+1}$ one can show that, for $\ell>0$,
\begin{align}
	\{L_1(t),Q_{2\ell+1}(t)\} &= -2\ell \int\d^2x\d^2y {\cal H}(x,y) (\partial_{x_1}+i\partial_{x_2})^{2\ell+2} G_{\ell+1}(\vx-\vy).
\end{align}
Changing now $\ell\to\ell-1$ one has
\begin{align}
	\{L_1(t),Q_{2\ell-1}(t)\} &= -2(\ell-1) \int\d^2x\d^2y {\cal H}(x,y) (\partial_{x_1}+i\partial_{x_2})^{2\ell} G_{\ell}(\vx-\vy)\nonumber\\
	&= -2(\ell-1) Q_{2\ell}(t) = \big(1-(2\ell-1)\big) Q_{2l}(t),\quad \ell >1,
\label{B2l+1}
\end{align}
which is the correct relation, and which again can be extended to the Poincar\'e $\ell=1$ case.

In order to get all the relations of the BMS algebra, an extra pair of brackets must be computed. The final results are
\begin{align}
	\{L_1(t),Q_{-2\ell}(t)\} &= (2\ell+1) Q_{-2\ell+1}(t) = (1-(-2\ell))Q_{-2\ell+1}(t),\label{Bm2l}\\
	\{L_1(t),Q_{-(2\ell+1)}(t)\} &= (2\ell+2) Q_{-2\ell}(t) = (1-(-(2\ell+1)))Q_{-2\ell}(t).\label{Bm2l+1}
\end{align}
In these cases the computations involve slightly different manipulations but always using (\ref{iden3}).

The brackets involving $L_{-1}$ can be computed from (\ref{B2l}), (\ref{B2l+1}), (\ref{Bm2l}) and (\ref{Bm2l+1}) by complex conjugation and using $L_1(t)=-L_{-1}^*(t)$, and one gets
\begin{align}
	\{L_{-1}(t),Q_{-2\ell}(t)\} &= \big(-1+2\ell\big) Q_{-(2\ell+1)}(t) =  \big(-1-(-2\ell)\big) Q_{-(2\ell+1)}(t), \label{Mm2l}\\
    \{L_{-1}(t),Q_{2\ell}(t)\} &= -\big(2\ell+1\big) Q_{2\ell-1}(t) =  \big(-1-2\ell)\big) Q_{2\ell-1}(t), \label{M2l}\\
    \{L_{-1}(t),Q_{2\ell+1}(t)\} &= -\big(2\ell+2\big) Q_{2\ell}(t) =  \big(-1-(2\ell+1)\big) Q_{2\ell}(t), \label{M2l+1}\\
    \{L_{-1}(t),Q_{-2\ell-1}(t)\} &= 2\ell Q_{-2(\ell+1)}(t) =  \big(-1-(-2\ell-1)\big) Q_{-2(\ell+1)}(t), \label{Mm2l+1}.
\end{align}
This completes the proof that the charges defined by (\ref{eq:charge_st_polyharmonic_2l}), (\ref{eq:charge_st_polyharmonic_2l+1}) and (\ref{ellneg}) provide, together with the Lorentz generators, a realization of the $2+1$ BMS algebra.

\section{The BMS algebra in configuration space}
\label{BMS-CS}
In the previous section we have shown that the BMS algebra is obtained in phase space using the expression of the supertranslation and Lorentz charges in terms of the fields $\phi$, $\pi$.  We will discuss now the transformations in configuration space, using $\phi$ and $\dot{\phi}$ as independent fields. The result is that one obtains a BMS algebra of transformations modulo trivial symmetry transformations, given by skew-symmetric combinations of the equations of motion of $\phi$.

In order to show this, let us consider the specific case of the Lorentz transformation $\delta_1^B$ associated to $L_1$ and the transformation $\delta_{2\ell}$  given by the supertranslation charge $Q_{2\ell}$, $\ell\geq 0$. In configuration space we must substitute $\dot{\phi}$ for $\pi$, and the transformations are
 
	\begin{align}
		 \delta^B_{1} \phi(x) &= t(\partial_{x_1} + i \partial_{x_2})\phi(x) + (x_1 + i x_2) \dot{\phi}(x), \\
		 \delta_{{2\ell}}  \phi(x) & = \int\d^2y \, \dot\phi(y) (\partial_{x_1}+i\partial_{x_2})^{2\ell} G_\ell(\vx-\vy).
 \end{align}
Since these are functional variations, the transformation of $\dot\phi$ is obtained by derivation, and one gets
	\begin{align}
	\delta^B_{1} \dot\phi(x) &= (\partial_{x_1} + i \partial_{x_2})\phi(x)+ t(\partial_{x_1} + i \partial_{x_2})\dot\phi(x)  + (x_1 + i x_2) \ddot{\phi}(x), \\
	\delta_{{2\ell}}  \dot\phi(x) & = \int\d^2y \, \ddot\phi(y) (\partial_{x_1}+i\partial_{x_2})^{2\ell} G_\ell(\vx-\vy).
\end{align}

Now we can compute the compositions of transformations
	\begin{align}
		\delta_{{2\ell}} \delta^B_{1} \phi(x) & = t(\partial_{x_1} + i \partial_{x_2}) \delta_{2\ell}\phi(x) + (x_1 + i x_2) \delta_{{2\ell}} \dot{\phi}(x) \nonumber\\
		& \!\!\!\!\!\!\!\!= t \int d^2y\, \dot{\phi}(y) (\partial_{x_1} + i \partial_{x_2})^{2\ell+1} G_\ell(\vec{x}-\vec{y}) 
		 + (x_1 + i x_2) \int d^2y\, \ddot{\phi}(y) (\partial_{x_1} + i \partial_{x_2})^{2\ell} G_\ell(\vec{x}-\vec{y}),
		 \label{eq:CE1}
	\end{align}
	and
	\begin{align}
		\delta^B_{1} \delta_{{2\ell}} \phi(x) & =  
		 \int d^2y\, \delta^B_{1}  \dot{\phi}(y) (\partial_{x_1} + i \partial_{x_2})^{2\ell} G_\ell(\vec{x}-\vec{y}) \nonumber \\
		& = \int d^2y\, (\partial_{y_1} + i \partial_{y_2})\phi(y) (\partial_{x_1} + i \partial_{x_2})^{2\ell} G_\ell(\vec{x}-\vec{y}) \nonumber\\
		 &+ t\int d^2y\, (\partial_{y_1} + i \partial_{y_2})\dot{\phi}(y) (\partial_{x_1} + i \partial_{x_2})^{2\ell} G_\ell(\vec{x}-\vec{y})\nonumber\\
		&+ \int d^2y\, (y_1 + i y_2) \ddot{\phi}(y) (\partial_{x_1} + i \partial_{x_2})^{2\ell} G_\ell(\vec{x}-\vec{y}).
		\label{eq:CE2}
	\end{align}
The first term in (\ref{eq:CE2}),
$$
\int d^2y\, (\partial_{y_1} + i \partial_{y_2})\phi(y) (\partial_{x_1} + i \partial_{x_2})^{2\ell} G_\ell(\vec{x}-\vec{y}) = \int d^2y\, \phi(y) (\partial_{x_1} + i \partial_{x_2})^{2\ell+1} G_\ell(\vec{x}-\vec{y})
$$
is just $-\delta_{2\ell+1}\phi(x)$ and, assembling the remaining terms, the commutator of the two transformations turns out to be
	\begin{align}
		 [\delta^B_{1},\delta_{{2\ell}}] \phi(x) &=  -\delta_{{2\ell+1}} \phi(x) 
		- \int d^2y\, \big((x_1-y_1) +  i (x_2 - y_2)\big)  \ddot{\phi}(y) (\partial_{x_1} + i \partial_{x_2})^{2\ell} G_\ell(\vec{x}-\vec{y}).
\end{align}
Now we add and subtract $\vN^2\phi$ and obtain 
\begin{eqnarray}
\lefteqn{[\delta^B_{1},\delta_{{2\ell}}]  \phi(x) =  -\delta_{{2\ell+1}} \phi(x) -
\int\d^2y\, \big( (x_1-y_1) + i (x_2-y_2)  \big)\vN^2_y\phi(y) (\partial_{x_1}+i\partial_{x_2})^{2\ell} G_\ell(\vx-\vy)}\nonumber\\
&-& 	\int\d^2y\, \big( (x_1-y_1) + i (x_2-y_2)  \big)\big(\ddot\phi(y)-\vN^2_y\phi(y)\big) (\partial_{x_1}+i\partial_{x_2})^{2\ell}G_\ell(\vx-\vy)\nonumber\\
&=& -\delta_{{2\ell+1}} \phi(x)  - 	\int\d^2y\, \big( (x_1-y_1) + i (x_2-y_2)	\phi(y) (\partial_{x_1}+i\partial_{x_2})^{2\ell}(\partial_{x_1}-i\partial_{x_2}) G_\ell(\vx-\vy)\nonumber\\
&-& 2\int\dif^2y\, \phi(y) (\partial_{x_1}+i\partial_{x_2})^{2\ell+1}G_\ell(\vx-\vy) 
-\int\d^2y\,  F_\ell(x,y) \big(\ddot\phi(y)-\vN^2_y\phi(y)\big),
\label{eq:CE3}
\end{eqnarray}
where we have integrated twice by parts the term $\vN^2\phi$ and defined
\begin{equation}
	F_\ell(x,y)= \big( (x_1-y_1) + i (x_2-y_2)  \big)(\partial_{x_1}+i\partial_{x_2})^{2\ell} G_\ell(\vx-\vy).
\end{equation}
The third term in (\ref{eq:CE3}) is $2\delta_{2\ell+1}\phi(x)$, while the second term, following the same steps that led to (\ref{eq3.15}), becomes $2(\ell-1)\delta_{2\ell+1}\phi(x)$. Putting everything together one has
\begin{align}
	[\delta^B_{1},\delta_{{2\ell}}]  \phi(x) &= (2\ell-1)\delta_{2\ell+1}\phi(x) - \int\d^2y\,  F_\ell(x,y) \big(\ddot\phi(y)-\vN^2_y\phi(y)\big).
	\label{eq:CE4}
\end{align}
Taking into account that, for any transformations generated by charges $A$, $B$, one has that $[\delta_A,\delta_B]\phi=-\delta_{\{A,B\}}\phi$, the first term in (\ref{eq:CE4}) is the one expected from the BMS algebra and, because $F_\ell(x,y)=-F_\ell(y,x)$, the extra term is a skew-symmetric linear combination of the equations of motion,
\begin{equation}
	\delta_{\ell,\text{trivial}}\phi(x) =  \int\d^2y\,  F_\ell(x,y) \big(\ddot\phi(y)-\vN^2_y\phi(y)\big),
\end{equation}
which is a trivial symmetry transformation of any system.
Notice that, for $\ell=0$, $F_\ell(x,y)=0$ and, as it must be, the extra term is not present for the standard commutator of a time translation and a Lorentz boost.

 Similar results are obtained for the other commutators of transformations, and hence the algebra closes on-shell in a consistent way.


\section{Discussion and outlook}
\label{diss}
We have obtained an explicit expression for the BMS supertranslation charges of free massless scalar real scalar field in $2+1$ space-time, in terms of the Green functions of the polyharmonic operator.

We work first in phase space, and discuss the asymptotic behaviour of the fields that ensures the existence of the charges, as well as that of the symplectic form associated to the Poisson brackets.

The conservation of the charges only depends on general symmetry properties of the involved functions, but the commutative character of the algebra satisfied by these charges relies on a convolution property of the Green functions. Finally, the correct algebra with the Lorentz generators is obtained using more specific properties of the polyharmonic Green functions.

We also discuss the closure of the transformations in configuration space, and it turns out that the correct algebra is obtained modulo transformations given by skew-symmetric combinations of the equations of motion, which are trivial symmetry transformations of any system. 

The form of the supertranslation charges presented in this paper, in terms of Green functions of an appropriate operator, opens the way to the generalization to other cases and/or dimensions. For instance, one could consider the extension of the results to the $3+1$ space-time dimension case, or to the case of the generators of superrotations which were also constructed in \cite{Batlle:2017llu}.

Another subject worth of study is whether the supertranslation transformations of the field $\phi$ can be interpreted as a base transformation of the coordinates in Minkowski space, or if additional coordinates, each associated to a supertranslation, must be introduced. 
%

\section{Acknowledgements} 
We acknowledge discussions with Marc Henneaux and Axel Kleinschmidt.

The work of CB is partially supported by Project PID2021-126001OB-C31 funded by MCIN/ AEI/ 10.13039/ 501100011033/ ERDF, EU.  JG has been supported in part by MINECO FPA2016-76005-C2-1-P and PID2019- 105614GB-C21 and from the State Agency for Research of the Spanish Ministry of Science and Innovation through the Unit of Excellence Maria de Maeztu 2020-2023 award to the Institute of Cosmos Sciences (CEX2019-000918-M).

\appendix

\section{Asymptotic behaviour of the polyharmonic Green functions}
\label{appG}
The polyharmonic Green function $G_\ell$ has the form
\begin{equation}
G_\ell(x) = A_\ell^{(0)} \mvx^{2\ell-2} \log\mvx + B_\ell^{(0)}\mvx^{2\ell-2},
\end{equation}
where the constants $A_\ell^{(0)}$ and $B_\ell^{(0)}$ can be read from (\ref{eq:fundamental}). Successive applications of $\partial_{x_1}+i\partial_{x_2}$ yield expressions of the same form, with decreasing powers of $\mvx$, until one reaches
\begin{equation}
(\partial_{x_1}+i\partial_{x_2})^{\ell-1} G_\ell(\vx) = (x_1+ix_2)^{\ell-1} \left(  A_\ell^{(\ell-1)} \log\mvx + B_\ell^{(\ell-1)}      \right).
\end{equation}
From this point the derivatives cease to content the $\log$ term and one can see that the derivative of order $2\ell$ is of the form
\begin{equation}
(\partial_{x_1}+i\partial_{x_2})^{2\ell} G_\ell(\vx)  = C^{(2\ell)} (x_1+i x_2)^{2\ell} \frac{1}{\mvx^{2\ell+2}},
\label{d2lG}
\end{equation}
with a constant $C^{(2\ell)}$. This is a rational function of $x_1$, $x_2$ with asymptotic behaviour
\begin{equation}
(\partial_{x_1}+i\partial_{x_2})^{2\ell} G_\ell(\vx)  \sim \frac{1}{\mvx^2}\quad \forall\ell\geq 1 \quad \text{for $\mvx\to\infty$},
\label{ABd2lG}
\end{equation}
which is independent of $\ell\geq 1$.
This allows us to study the conditions that must be imposed on the fields so that the supertranslation charges are finite. A general supertraslation charge $Q_\ell$, $\ell\geq1$, has integrals of the form
\begin{equation}
Q_\ell(t) = \int\d^2x\d^2y F(t,\vx)G(t,\vy) (\partial_{x_1}+i\partial_{x_2})^{2\ell} G_\ell(\vx-\vy) 
\end{equation}
 where  $F$ and $G$ are either $\pi$ or first order derivatives of $\phi$.  Performing a change of variables
 $\vx=\vy+\vr$ and using the asymptotic behaviour (\ref{ABd2lG}), the existence of the charge reduces to the existence of the integral
 \begin{equation}
 \int\dif^2r\dif^2 y F(\vy+\vr)G(\vy) \frac{1}{r^2}.
 \label{asimp1}
 \end{equation}
 Let us assume now that the fields $F$ and $G$ behave, for large argument, as
 \begin{equation}
 F(\vy+\vr) \sim \frac{\bar{F}}{|\vy+\vr|^\alpha}, \quad G(\vy) \sim \frac{\bar{G}}{|\vy|^\beta} ,
 \end{equation}
 with $\bar{F}$, $\bar{G}$ depending on the angular variable and time. This leads to the study of the integral
   \begin{equation}
  \int\dif^2\Omega\, \bar{F}\bar{G} \int\, r\dif r\, y\dif y\,  \frac{1}{|\vy+\vr|^\alpha|\vy|^\beta r^2},
  \label{asimp2}
  \end{equation}
where $\dif^2\Omega$ is the angular measure. Performing a change to polar coordinates in $\R^2_+$ for the $\dif r\dif y$ measure, one finally gets
 \begin{equation}
\int\dif^3\Omega\, \bar{F}\bar{G} \int\, \rho\dif \rho\,  \frac{1}{\rho^{\alpha+\beta}},
\label{asimp3}
\end{equation}
with $\dif^3\Omega$ including the additional integration over the angular coordinate of the polar change of variables. For this integral to converge it is necessary that
\begin{equation}
\alpha+\beta > 2.
\end{equation}
Considering the forms of $F$ and $G$   for the different supertranslation charges, one concludes that  the asymptotic behaviour of the fields which guarantees the existence of all the $Q_\ell$ is the one given in
(\ref{asphi}) and (\ref{aspi}).

\section{Brackets between the supertranslation charges}
\label{appA}
Consider two arbitrary supertranslation charges
\begin{align}
Q_\ell(t) &= \int \d^2x \d^2y\,  (f_\ell (\vec{x}-\vec{y})\pi(x) \phi(y) + \dfrac{1}{2} g_\ell (\vec{x}-\vec{y}) \pi(x) \pi(y) 
-\dfrac{1}{2} h_\ell (\vec{x}-\vec{y}) \phi(x) \phi(y)),\\
Q_m(t) &= \int \d^2z \d^2w\,  (f_m (\vec{z}-\vec{w})\pi(z) \phi(w) + \dfrac{1}{2} g_m (\vec{z}-\vec{w}) \pi(z) \pi(w) 
-\dfrac{1}{2} h_m (\vec{z}-\vec{w}) \phi(z) \phi(w)).
\end{align}
Using standard Poisson brackets, one gets
\begin{align}
	\{ Q_\ell(t),Q_m(t)\} &= \int\d^2x\d^2y\d^2z\, \Big(
	 f_\ell(\vx-\vy)f_m(\vy-\vz) \pi(x)\phi(z) - f_\ell(\vx-\vy)f_m(\vz-\vx)\phi(y)\pi(z)
	 \nonumber\\
	&+f_\ell(\vx-\vy)g_m(\vy-\vz)\pi(x)\pi(z)+f_\ell(\vx-\vy)h_m(\vx-\vz)\phi(y)\phi(z) \nonumber\\
	&- g_\ell(\vx-\vy)f_m(\vz-\vx)\pi(y)\pi(z) +g_\ell(\vx-\vy)h_m(\vx-\vz)\pi(y)\phi(z)
	\nonumber\\
	&-h_\ell(\vx-\vy)f_m(\vx-\vz)\phi(y)\phi(z) 
	- h_\ell(\vx-\vy)g_m(\vx-\vz)\phi(y)\pi(z)
	\Big).
	\label{pbQQ}
\end{align}

Let us consider first the case $m=0$, that is $Q_m(t)=Q_0(t)=H(t)$, for which $f_0(\vx-\vy)=0$, $g_0(\vx-\vy)=\delta(\vx-\vy)$ and $h_0(\vx-\vy)=\vN^2_x\delta(\vx-\vy)$. After integration by parts one gets 
\begin{align}
	\dot{Q}_\ell(t) &= 	\{ Q_\ell(t),H(t)\} = \int\d^2x\d^2y\, \Big(
	f_\ell(\vx-\vy)\pi(x)\pi(y)+ \vN^2_xf_\ell(\vx-\vy)\phi(y)\phi(x)\nonumber\\
	&+ \vN^2_x g_\ell(\vx-\vy)\pi(y)\phi(x) - h_\ell(\vx-\vy)\phi(y)\pi(x)
	\Big).
\end{align}
The first two terms are zero, each by itself, due to the skew-symmetry of $f_\ell$ and its even-order derivatives, while the two last terms cancel each other after using $\vN^2(x)g_\ell(\vx-\vy)=h_\ell(\vx-\vy)$ and the symmetry of $h_\ell$.
This shows that the supertraslation charges are conserved by virtue of the symmetry properties of $f_\ell$, $g_\ell$ and $h_\ell$, and the relation between $g_\ell$ and $h_\ell$, without using the explicit form of these functions in terms of the polyharmonic Green functions.

For general $\ell$ and $m$ one must consider the different cases separately.
\begin{enumerate}
	\item $\ell$ and $m$ odd. In this case $g_\ell=h_\ell=g_m=h_m=0$ and, after renaming the variables of integration in the first non-zero contribution,
\begin{align}
	\{ Q_\ell(t),Q_m(t)\} &= \int\d^2x\d^2y\d^2z\, \Big(
	f_\ell(\vz-\vx)f_m(\vx-\vy) - f_\ell(\vx-\vy)f_m(\vz-\vx)\Big)\phi(y)\pi(z).
	\label{pbQQoo}
\end{align}	
\item $\ell$ even and $m$ odd. Now $f_\ell=0$ and $g_m=h_m=0$, and the result can be written as
\begin{align}
	\{ Q_\ell(t),Q_m(t)\} &= -\int\d^2x\d^2y\d^2z\,
	 g_\ell(\vx-\vy)f_m(\vz-\vx)\pi(y)\pi(z) \nonumber\\
	&-\int\d^2x\d^2y\d^2z\, h_\ell(\vx-\vy)f_m(\vx-\vz)\phi(y)\phi(z).
	\label{pbQQeo}
\end{align}
\item $\ell$ and $m$ even. We have $f_\ell=f_m=0$ and (\ref{pbQQ}) boils down to
\begin{align}
	\{ Q_\ell(t),Q_m(t)\} &= \int\d^2x\d^2y\d^2z\,\Big(
	g_\ell(\vx-\vz)h_m(\vx-\vy)
	- h_\ell(\vx-\vy)g_m(\vx-\vz)\Big)\phi(y)\pi(z).
	\label{pbQQee}
\end{align}
\end{enumerate}

The elementary symmetry properties used up to now are not enough to show that the above expressions are actually zero. In order to do so, one must use the fact that the functions $f_\ell$, $g_\ell$ and $h_\ell$ can be written in terms of Green functions that obey  the convolution property (\ref{convGG}).

Let us prove, for instance, that the first term in (\ref{pbQQeo}) is zero. Changing $\ell\to 2\ell$ and $m\to 2m+1$ one has, assuming $l>0$, $m>0$,   
\begin{eqnarray*}
\lefteqn{-\int\d^2x\d^2y\d^2z\,
	g_\ell(\vx-\vy)f_m(\vz-\vx)\pi(y)\pi(z) \rightarrow}\\
&-&\int\d^2x\d^2y\d^2z\, (\partial_{x_1}+i\partial_{x_2})^{2\ell}G_\ell(\vx-\vy)
 (\partial_{x_1}+i\partial_{x_2})^{2m+1}G_m(\vx-\vz) \pi(y)\pi(z) \\
&=& -\int\d^2x\d^2y\d^2z\, G_\ell(\vx-\vy)
G_m(\vx-\vz) (\partial_{y_1}+i\partial_{y_2})^{2\ell}\pi(y)(\partial_{z_1}+i\partial_{z_2})^{2m+1}\pi(z) \\ 
&\stackrel{(\ref{convGG})}{=}& -\int\d^2y\d^2z G_{\ell+m}(\vy-\vz) 
(\partial_{y_1}+i\partial_{y_2})^{2\ell}\pi(y)(\partial_{z_1}+i\partial_{z_2})^{2m+1}\pi(z) \\
&=& -\int\d^2y\d^2z (\partial_{y_1}+i\partial_{y_2})^{2\ell+2m+1}G_{\ell+m}(\vy-\vz) 
\pi(y)\pi(z)=0
\end{eqnarray*}	
due to the skew-symmetry of $(\partial_{y_1}+i\partial_{y_2})^{2\ell+2m+1}G_{\ell+m}(\vy-\vz)$, and the second term of  (\ref{pbQQeo}) can also be shown to be  zero using the same manipulations. Notice that the same reasoning can be used for $\ell$ and/or $m$ negative, since this amounts to change some $\partial_{x_1}+i\partial_{x_2}$ to $\partial_{x_1}-i\partial_{x_2}$ and the result, which only depends on the number of derivatives, is the same.

Using the same techniques and the convolution property, the two terms which appear in (\ref{pbQQoo}) or (\ref{pbQQee}) can be shown to be the same and hence that the corresponding brackets are zero. This completes the proof that the supertraslation charges yield a commutative algebra under the Poisson brackets.

\section{Detailed computation of some Poisson brackets}
\label{appP}

Consider first
\begin{eqnarray*}
	\lefteqn{\{L_0(t),Q_{2\ell}(t)\}= } \\
	&= -\frac{1}{2i}\int\d^2 x\d^2y\d^2z  \left\{
	\pi(x) (x_1\partial_{x_2}\phi(x)-x_2\partial_{x_1}\phi(x)), {\cal H}(y,z)
	\right\} (\partial_{y_1}+i\partial_{y_2})^{2\ell} G_\ell(\vy-\vz).
\end{eqnarray*}

Using that even-order derivatives of an even-symmetric function are even, and the equal-time Poisson brackets $\{\phi(x),\pi(y) \}=\delta(\vx-\vy)$, the above boils down to
\begin{align*}
	\{L_0(t),Q_{2\ell}(t)\} &= -\frac{1}{2i}\int\d^2 x\d^2y\d^2z \Big(
	(x_1\partial_{x_2}\phi(x)-x_2\partial_{x_1}\phi(x)) \vN\phi(z)\cdot\vN_y(-\delta(\vx-\vy))\\
	& + \pi(x)\pi(z) (x_1\partial_{x_2}-x-2\partial_{x_1})\delta(\vx-\vy)\Big)  (\partial_{y_1}+i\partial_{y_2})^{2\ell} G_\ell(\vy-\vz)\\
	& = -\frac{1}{2i}\int\d^2 x\d^2z \Big({2\cal{H}}(x,z)\Big) (x_1\partial_{x_2}-x_2\partial_{x_1})(\partial_{x_1}+i\partial_{x_2})^{2\ell} G_\ell(\vx-\vz),
\end{align*}
where several integrations by parts, assuming the appropriate asymptotic behaviour for the fields, have been performed, and the relation $\vN_xG_{\ell}(\vx-\vy)=-\vN_yG_{\ell}(\vx-\vy)$ has been used. Next, we use the commutator
\begin{equation}
	\big[
	x_1\partial_{x_2} - x_2\partial_{x_1}, (\partial_{x_1}+i\partial_{x_2})^n
	\big] = i n (\partial_{x_1}+i\partial_{x_2})^n, \quad n=0,1,\ldots 
\end{equation}
to write the above as
\begin{align*}
	\{L_0(t),Q_{2\ell}(t)\} &=	 -\frac{1}{i}\int\d^2 x\d^2z {\cal H}(x,z) \Big(
	(\partial_{x_1}+i\partial_{x_2})^{2\ell} (x_1\partial_{x_2}-x_2\partial_{x_1})\\
	&\quad  + i\ 2\ell\ (\partial_{x_1}+i\partial_{x_2})^{2\ell} 
	\Big) G_\ell(\vx-\vz).
\end{align*}
One has that
\begin{equation}
	(x_1\partial_{x_2}-x_2\partial_{x_1}) G_\ell(\vx-\vz) = G'_{\ell}(|\vx-\vz|) \frac{-x_1z_2+x_2 z_1}{|\vx-\vz|}
\end{equation}
is and odd function under $\vx\leftrightarrow\vz$, and hence its derivatives of even order are also odd. The product with ${\cal H}(x,z)$ is also odd and the term vanishes under integration in $x$ and $z$.  Thus one is left only with the term from the commutator and
\begin{align}
	\{L_0(t),Q_{2\ell}(t)\} &= -2\ell \int\d^2 x\d^2z 	 {\cal H}(x,z) (\partial_{x_1}+i\partial_{x_2})^{2\ell} G_\ell(\vx-\vz)\nonumber\\
	&= -2\ell Q_{2\ell}(t),\quad \ell\geq 0.
	\label{appPL2l}
\end{align}

Let us compute now the Poisson bracket of the boost generator $L_1(t)$ with an even-order supertranslation charge,
\begin{eqnarray*}
	\lefteqn{\{L_1(t),Q_{2\ell}(t)\}= } \\
	&=& \int\d^2 x\d^2y\d^2z  \Big\{
	t\pi(x) (\partial_{x_1}+i\partial_{x_2}) \phi(x) + (x_1+ix_2){\cal H}(x), {\cal H}(y,z)
	\Big\} (\partial_{y_1}+i\partial_{y_2})^{2\ell} G_\ell(\vy-\vz).
\end{eqnarray*}

After computing the Poisson brackets and using integration by parts and the symmetry of $G_\ell$ and its derivatives, the terms containing $t$ can be written as
$$
2 t \int\d^2x\d^2z {\cal H}(x,z) (\partial_{x_1}+i\partial_{x_2})G_\ell(\vx-\vz),
$$
which is zero due to the skew-symmetry of $(\partial_{x_1}+i\partial_{x_2})G_\ell(\vx-\vz)$ under $\vx\leftrightarrow\vz$. Performing the same manipulations, the remaining terms can be written as
\begin{align}
	\{L_1(t),Q_{2\ell}(t)\} &= \int\d^2x\d^2z \Big(
	(x_1+ix_2) \pi(x) \vN\phi(z) (\partial_{x_1}+i\partial_{x_2})^{2\ell}\cdot\vN_xG_\ell(\vx-\vz)\nonumber\\
	& \quad  +  (x_1+ix_2) \vN\phi(x)\pi(z)  (\partial_{x_1}+i\partial_{x_2})^{2\ell}\cdot\vN_xG_\ell(\vx-\vz)
	\Big).
\end{align}
The $\vN\phi(z)$ in the first term can be integrated by parts, yielding one term, while the integration by parts of $\vN\phi(x)$ yields two. After a change of variables and using 
$(\partial_{z_1}+i\partial_{z_2})^{2\ell}\vN_z^2G_l(\vx-\vz)=(\partial_{x_1}+i\partial_{x_2})^{2\ell}\vN_x^2G_l(\vx-\vz)$, the two terms that are similar can be combined, and the result is
\begin{align}
	\{L_1(t),Q_{2\ell}(t)\} &= \int\d^2x\d^2z \Big(
	((x_1-z_1)+i(x_2-z_2)) \pi(x) \phi(z) (\partial_{x_1}+i\partial_{x_2})^{2\ell}\cdot\vN_x^2 G_\ell(\vx-\vz)\nonumber\\
	& \quad  - \phi(x)\pi(z)  (\partial_{x_1}+i\partial_{x_2})^{2\ell} (\partial_{x_1}+i\partial_{x_2})G_\ell(\vx-\vz)
	\Big).
\end{align}
The second term is just $-Q_{2\ell+1}(t)$ and the integrand in the first one can be re-written, using that $[x_1+ix_2,\partial_{x_1}+i\partial_{x_2}]=0$ and $\vN^2_x=(\partial_{x_1}+i\partial_{x_2})(\partial_{x_1}-i\partial_{x_2})$, as
\begin{eqnarray}
	\lefteqn{\{L_1(t),Q_{2\ell}(t)\} =-Q_{2\ell+1}(t)}\nonumber\\
	&\!+ \int\d^2x\d^2z \Big(
	\pi(x) \phi(z) (\partial_{x_1}+i\partial_{x_2})^{2\ell+1}
	((x_1-z_1)+i(x_2-z_2)) (\partial_{x_1}-i\partial_{x_2})  G_\ell(\vx-\vz)
	\Big).
	\label{eq3.15}
\end{eqnarray}
Since we are considering $\ell\geq 1$ (the case $\ell=0$ correspond to the standard Poincaré algebra) and $2\ell+1> 2(\ell-1)$, we can use relation (\ref{iden3}) to rewrite (\ref{eq3.15}) as
\begin{align}
	\{L_1(t),Q_{2\ell}(t)\} &= -Q_{2\ell+1}(t)  + 2(\ell-1) \int\d^2x\d^2z \pi(x) \phi(z) (\partial_{x_1}+i\partial_{x_2})^{2\ell+1} G_\ell(\vx-\vz)\nonumber\\
	&= -Q_{2\ell+1}(t)  - 2(\ell-1) \int\d^2x\d^2z  \phi(x) \pi(z) (\partial_{x_1}+i\partial_{x_2})^{2\ell+1} G_\ell(\vx-\vz)\nonumber\\
	&=  -Q_{2\ell+1}(t) - 2(\ell-1) Q_{2\ell+1}(t)\nonumber\\
	&= (1-2\ell) Q_{2\ell+1}(t).
	\label{appPB2l}
\end{align}

\section{Some identities satisfied by the polyharmonic Green functions}
\label{appB}
From (\ref{eq:fundamental}) one has
\begin{align}
	(\partial_{x_1}-i\partial_{x_2}) G_\ell(\vx-\vy) &= \frac{|\vx-\vy|^{2(\ell-2)}}{(\ell-1)[(\ell-2)!]^2 2^{2\ell-2}\pi} ((x_1-y_1)-i (x_2-y_2))
	\big(\log|\vx-\vy| - H_{\ell-1}\big)\nonumber\\
	&+ \frac{|\vx-\vy|^{2(\ell-2)}}{[(\ell-1)!]^2 2^{2\ell-1}\pi} ((x_1-y_1)-i (x_2-y_2)).
\end{align}
Assuming $\ell>1$ and using $H_{\ell-1} = H_{\ell-2} + 1/(\ell-1)$ and re-arranging terms, this can be rewritten as
\begin{align}
(\partial_{x_1}-i\partial_{x_2}) G_\ell(\vx-\vy) &= \frac{1}{2(\ell-1)} ((x_1-y_1)-i (x_2-y_2)) G_{\ell-1} (\vx-\vy) \nonumber\\
 & - \frac{|\vx-\vy|^{2(\ell-2)}}{[(\ell-1)!]^2 2^{2\ell-1}\pi} ((x_1-y_1)-i (x_2-y_2)),
\label{iden1}
\end{align}
which is a recurrence relation valid for $\ell>1$.  
Multiplying by $(x_1-y_1) + i (x_2-y_2)$ one gets
\begin{align}
	((x_1-y_1) + i (x_2-y_2))(\partial_{x_1}-i\partial_{x_2}) G_\ell(\vx-\vy) &= \frac{1}{2(\ell-1)} |\vx-\vy|^2 G_{\ell-1} (\vx-\vy) \nonumber\\
	& - \frac{|\vx-\vy|^{2(\ell-1)}}{[(\ell-1)!]^2 2^{2\ell-1}\pi},
\end{align}
which, except for polynomial terms, has the functional dependence of $G_\ell$. Indeed, using again the relation between $H_{\ell-1}$ and $H_{\ell-2}$, one obtains
\begin{align}
	((x_1-y_1) + i (x_2-y_2))(\partial_{x_1}-i\partial_{x_2}) G_\ell(\vx-\vy) = 2(\ell-1) G_{\ell} (\vx-\vy)  + \frac{|\vx-\vy|^{2(\ell-1)}}{[(\ell-1)!]^2 2^{2\ell-1}\pi}.
	\label{iden2}
\end{align}
Although (\ref{iden2}) has been obtained under the assumption that $\ell>1$ it can be checked by direct computation that it is also valid for $\ell=1$.

In the computations in Section \ref{BMS-PS} the left-hand side of this identity appears with derivatives $\partial_{x_1}+i\partial_{x_2}$ acting on it. Since the second term in (\ref{iden2})
is a polynomial of order $2(\ell-1)$ in the components of $\vx$, it turns out that, for $\ell\geq 1$,
  \begin{equation}
   (\partial_{x_1}+i\partial_{x_2})^n((x_1-y_1) + i (x_2-y_2))(\partial_{x_1}-i\partial_{x_2}) G_\ell(\vx-\vy) 
  	 = 2(\ell-1)(\partial_{x_1}+i\partial_{x_2})^n G_{\ell} (\vx-\vy)  
  	\label{iden3}
  \end{equation}
for $n>2(\ell-1)$.

\section{Convolution property of the polyharmonic Green functions}
\label{appC}
One has
\begin{align}
	\lefteqn{\vN_y^{2(\ell+m)} \int\d^2x\, G_\ell(\vy-\vx)G_m(\vz-\vx)}\nonumber\\
	&=  \int\d^2x\, \vN_y^{2(\ell+m)}G_\ell{\vy-\vx}G_m(\vz-\vx) 
	= \int\d^2x\, \vN_y^{2\ell} \vN_x^{2m}  G_\ell{\vy-\vx}G_m(\vz-\vx) \nonumber\\
	&= 	\int\d^2x\, \vN_y^{2\ell}   G_\ell{\vy-\vx}  \vN_x^{2m}G_m(\vz-\vx)
	= 	\int\d^2x\, \delta(\vy-\vx) \delta(\vz-\vx)\nonumber\\ 
	&= \delta(\vy-\vz).
\end{align}
Under standard regularity conditions, the homogeneous polyharmonic problem  has only the trivial solution \cite{Gazzola2010}, and the above computation proves (\ref{convGG}).

\bibliographystyle{JHEP}
\bibliography{bms.bib}{}

\end{document}